\documentclass[12pt]{article}
\usepackage{amsfonts,latexsym}
\def\bea{\begin{eqnarray}}
\def\eea{\end{eqnarray}} 
\def\nn{\nonumber} 

\begin{document}

\begin{center}
{\large Integrability and exact solution for coupled BCS systems 
associated with the $su(4)$ Lie algebra}\\
\vskip.3in
{ Xi-Wen Guan $^{\dag}$, Angela Foerster $^{\dag,\ddag \ddag}$, 
Jon Links $^{\dag \dag}$\\
and Huan-Qiang Zhou $^{\dag \dag}$}
\vskip.2in
$^{\dag}$Instituto de F\'{\i}sica da UFRGS,
                     Av.\ Bento Gon\c{c}alves, 9500,\\
                     Porto Alegre, 91501-970, Brasil\\
$^{\ddag \ddag}$ Institut f\"{u}r Theoretische Physik, der FU-Berlin,\\
 Arnimallee 14, Berlin, Germany\\
$^{\dag \dag}$Centre for Mathematical Physics, 
School of Physical Sciences, \\ The University of Queensland, 4072, Australia
\end{center}

\begin{abstract}
We introduce an integrable model for two coupled BCS systems through 
a solution of the Yang-Baxter equation associated with the Lie algebra
$su(4)$. By employing the algebraic Bethe ansatz, we  determine the
exact solution for the energy spectrum. An asymptotic analysis is
conducted to determine the leading terms in the ground state energy, the
gap and some one point correlation functions at zero temperature. 
\end{abstract}

PACS: 
71.24+q, 74.20Fg


\newpage
\section{Introduction}
\label{sec1}

The role of the Yang-Baxter equation in the study of quantum mechanical
models has a long and distinguished history. Notable examples are the 
$XYZ$ chain \cite{baxter}, the $t-J$ model at supersymmetric coupling 
\cite{tj} and the Hubbard model \cite{lw}, 
each of which is both integrable and exactly solvable as a result
of the formulation for each model through the Quantum Inverse Scattering
Method (QISM). The key concept of the QISM is the notion of mutually
commuting transfer matrices, the existence of which is a result of the
Yang-Baxter equation. 
In each of the above examples  
the Hamiltonian of the model is defined as the logarithmic derivative of
the transfer matrix, and by the nature of the construction this yields 
a model defined on a one-dimensional lattice with nearest neighbour
interactions, where it is possible for both integrability and solvability to
hold for a variety of boundary conditions. 

The application of the QISM however can be applied on a much more
general level. In the context of the present work, it is appropriate to
mention, for example,  
the work of Gaudin \cite{gaudin} in relation to the construction of
systems with long range interactions. Very closely related to Gaudin's 
Hamiltonians is the BCS model, the exact solution of which was quite
remarkably found in 1963 by Richardson \cite{r63}, while integrability was 
established much later by Cambiaggio et al. in 1997 \cite{crs97}. The exact
solution of the BCS model has come under close scrutiny in recent years
due to its application in the theory of metallic nanograins \cite{dr01}.
Specifically, the experiments of Ralph, Black and Tinkham \cite{rbt} have
shown that it is not valid to apply 
the BCS mean field theory for systems of nanoscale size. As a result,
one has to turn to the exact solution of Richardson in order to conduct
a reliable analysis. However, the approaches adopted in
\cite{r63,crs97} make no reference to the Yang-Baxter equation or the QISM,
(indeed QISM was not developed until many years after Richardson's
work),
and since historically the two facets of integrability and solvability 
have  been intimately linked in the
QISM framework, it was a natural question to ask whether the BCS model
could be recast through this technique. An affirmative answer was given
in \cite{zlmg,vp} with the surprising result that the $R$-matrix solution of
the Yang-Baxter equation which is needed in the construction of the BCS
Hamiltonian is one of the simplest known examples, being that associated
with the $su(2)$ algebra. Given that a great volume of literature exists devoted to
solutions of the Yang-Baxter equation associated with representations of
simple Lie algebras, there is a vast opportunity to investigate generalised
models. An important step towards this has already been achieved in
\cite{afs} where a connection has been established between Chern-Simons
theory and integrability of models associated with an arbitrary Lie
algebra, which is achieved through the Knizhnik-Zamolodchikov-Bernard
equations.   

Such generalised models can be interpreted as coupled BCS systems, at
least in the sense that every simple Lie algebra can be generated by a
system of simple roots which each form an $su(2)$ subalgebra. An
example of this was given in \cite{lzgm} where the Lie algebra
employed was $so(5)$.  In this instance, the model constructed
describes proton-proton and  neutron-neutron pairing 
as well as a coupling term
for the scattering of proton-neutron pairs.  Here we shall introduce a
model based on the $su(4)$ Lie algebra symmetry which can also be
interpreted as a nuclear system where there are now different types of
pairing interactions. The Hamiltonian takes the form of two BCS
systems which individually describe pairing interactions for the
protons and neutrons and the scattering of bound proton pairs-neutron
pairs, which is in contrast to the proton-neutron pairs of
\cite{lzgm}.  This interpretation is possible because the number
operator for each system provides a good quantum number; i.e. the
number operators are conserved. Therefore we can identify each BCS
system with a particular distinguishable particle, which in this case
are the protons and neutrons. It is worth remarking that this
situation is inherently different to the pairing models described in
\cite{jon1} based on higher spin representations of the $su(2)$
algebra. In these cases, the only good quantum number is the total
number of particles in the combined system.  There, individual
particle numbers are not conserved and thus the models can be
interpreted as describing a Josephson tunneling phenomena.

The paper is organized as follows. In Section 2 we present the
construction of the model through the QISM.  In section 3
the exact solution of the model is given by means of the algebraic
Bethe ansatz. An analysis of the asymptotic solutions of the Bethe
ansatz equations is presented in Section 4 , where the ground state
energy, the gap in the spectrum of elementary excitations, as well as
the derivation of some correlation functions in this asymptotic regime,
are presented. A summary of the main results can be found in Section 5.
\section{Coupled pairing Hamiltonian  and  integrability}
\label{sec2}
Let us begin by introducing the following Hamiltonian
\begin{eqnarray}
H &=&BCS(1)+BCS(2)
-g\sum_{j,k}^{\Omega}b_j^+(1)b_j^+(2)b_k(2)b_k(1)\nonumber\\
&&+g\sum_{j,k}^{\Omega}b_j^+(1)b_k(1)\left(n_j(2)-n_k(2)\right)^2\nonumber\\
& &
+g\sum_{j,k}^{\Omega}b_j^+(2)b_k(2)\left(n_j(1)-n_k(2)\right)^2,
 \label{Ham}
 \end{eqnarray}
 where 
 \begin{equation}
BCS(a)=\sum^{\Omega}_{j=1}2\epsilon_jn_j(a)-g\sum^{\Omega}_{j,k}b^+_j(a)b_k(a).
\end{equation}
Above the operators $b_j(a),\,b_j^{+}(a)$ are the annihilation and creation
operators for the hard-core bosons (or Cooper pairs) in system $a$, and
$j$ refers to the single particle energy level with energy $\epsilon_j$.  
We will assume that the values $\epsilon _j$ are distinct. 
Further, $g$ is a coupling strength constant for  
the scattering of Cooper pairs and 
 $n_j(a)=b_j^+(a)b_j(a)$, is the Cooper pair number
 operator. As in the case of the usual BCS system there is a blocking
 effect (e.g. see \cite{dr01}),  
 as there is no scattering of any unpaired states. For each
 level $j$ there are actually sixteen local states, but the nature of the
 Hamiltonian means that only on  a subspace spanned by four of these
 states, where there are no unpaired states, is the scattering non-trivial
 (see (\ref{states}) below). 
Hereafter we will restrict our analysis to this subspace. 

On this restricted subspace the operators
 $b_j^+(a)=c^{\dagger}_{j\uparrow}(a)c^{\dagger}_{j\downarrow}(a),
   ~b_j(a)=c_{j\downarrow}(a)c_{j\uparrow}(a)$,   
   where
   $c_{j\sigma},\,c^{\dagger}_{j\sigma},\,\sigma=\uparrow,\,\downarrow$,
   are the familiar fermion operators, 
   satisfy the hard-core boson relations
   \begin{eqnarray}
   & &(b^+_j(a))^2=0,~~~
   [b_j(a),b_k^+(b)]=\delta_{ab}\delta_{jk}(1-2b^+_j(a)b_j(a)),\nonumber\\
   & &
   [b_j(a),b_k(b)]=[b^+_j(a),b^+_k(b)]=0, ~~~\rm{for}~~k\neq j.\nonumber
   \end{eqnarray}

We can see from the Hamiltonian expression  that the
exchange interaction of Cooper pairs in one system depends on the
number of Cooper pairs in the other system. For example, if in 
system (1), the level $j$ is empty and the level $k$ is occupied by one
Cooper pair, just for certain configurations of system (2) it is
possible that this Cooper pair in  (1) scatters from level $k$ to
$j$. This means that the Hamiltonian (\ref{Ham}) presents naturally some
``selection rules'' for the scattering of states. We illustrate
these configurations to indicate the possible pair scatterings
$$
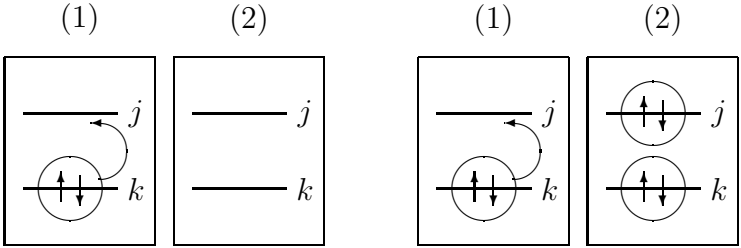
\unitlength=0.50mm
\begin{picture}(250.,75.)
\put(20.,60.){\makebox(0,0)[cc]{$(1)$}}
\put(0.,0.){\line(0,1){50.}}
\put(40.,0.){\line(0,1){50.}}
\put(0.,0.){\line(1,0){40.}}
\put(0.,50.){\line(1,0){40.}}
\put(5.,15.){\line(1,0){25.}}
\put(5.,35.){\line(1,0){25.}}
\put(35.,15.){\makebox(0,0)[cc]{$k$}}
\put(35.,35.){\makebox(0,0)[cc]{$j$}}
\put(17.5,15.){\oval(17.,17.)[l]}
\put(17.5,15.){\oval(17.,17.)[r]}
\put(15.,11.5){\vector(0,1){8.}}
\put(20.,18.5){\vector(0,-1){8.}}
\put(25,25.){\oval(15.,15.)[r]}
\put(23.,32.5){\vector(-1,0){0.1}}
\put(65.,60.){\makebox(0,0)[cc]{$(2)$}}
\put(45.,0.){\line(0,1){50.}}
\put(85.,0.){\line(0,1){50.}}
\put(45.,0.){\line(1,0){40.}}
\put(45.,50.){\line(1,0){40.}}
\put(50.,15.){\line(1,0){25.}}
\put(50.,35.){\line(1,0){25.}}
\put(80.,15.){\makebox(0,0)[cc]{$k$}}
\put(80.,35.){\makebox(0,0)[cc]{$j$}}
\put(130.,60.){\makebox(0,0)[cc]{$(1)$}}
\put(110.,0.){\line(0,1){50.}}
\put(150.,0.){\line(0,1){50.}}
\put(110.,0.){\line(1,0){40.}}
\put(110.,50.){\line(1,0){40.}}
\put(115.,15.){\line(1,0){25.}}
\put(115.,35.){\line(1,0){25.}}
\put(145.,15.){\makebox(0,0)[cc]{$k$}}
\put(145.,35.){\makebox(0,0)[cc]{$j$}}
\put(127.5,15.){\oval(17.,17.)[l]}
\put(127.5,15.){\oval(17.,17.)[r]}
\put(125.,11.5){\vector(0,1){8.}}
\put(130.,18.5){\vector(0,-1){8.}}
\put(135,25.){\oval(15.,15.)[r]}
\put(133.,32.5){\vector(-1,0){0.1}}
\put(175.,60.){\makebox(0,0)[cc]{$(2)$}}
\put(155.,0.){\line(0,1){50.}}
\put(195.,0.){\line(0,1){50.}}
\put(155.,0.){\line(1,0){40.}}
\put(155.,50.){\line(1,0){40.}}
\put(160.,15.){\line(1,0){25.}}
\put(160.,35.){\line(1,0){25.}}
\put(190.,15.){\makebox(0,0)[cc]{$k$}}
\put(190.,35.){\makebox(0,0)[cc]{$j$}}
\put(172.5,15.){\oval(17.,17.)[l]}
\put(172.5,15.){\oval(17.,17.)[r]}
\put(170.,11.5){\vector(0,1){8.}}
\put(175.,18.5){\vector(0,-1){8.}}
\put(172.5,35.){\oval(17.,17.)[l]}
\put(172.5,35.){\oval(17.,17.)[r]}
\put(170.,31.5){\vector(0,1){8.}}
\put(175.,38.5){\vector(0,-1){8.}}
\end{picture}
$$
In addition, the double-pair scattering terms of the form  
$$
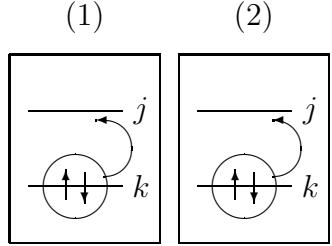
\unitlength=0.50mm
\begin{picture}(150.,75.)
\put(20.,60.){\makebox(0,0)[cc]{$(1)$}}
\put(0.,0.){\line(0,1){50.}}
\put(40.,0.){\line(0,1){50.}}
\put(0.,0.){\line(1,0){40.}}
\put(0.,50.){\line(1,0){40.}}
\put(5.,15.){\line(1,0){25.}}
\put(5.,35.){\line(1,0){25.}}
\put(35.,15.){\makebox(0,0)[cc]{$k$}}
\put(35.,35.){\makebox(0,0)[cc]{$j$}}
\put(17.5,15.){\oval(17.,17.)[l]}
\put(17.5,15.){\oval(17.,17.)[r]}
\put(15.,11.5){\vector(0,1){8.}}
\put(20.,18.5){\vector(0,-1){8.}}
\put(25,25.){\oval(15.,15.)[r]}
\put(23.,32.5){\vector(-1,0){0.1}}
\put(65.,60.){\makebox(0,0)[cc]{$(2)$}}
\put(45.,0.){\line(0,1){50.}}
\put(85.,0.){\line(0,1){50.}}
\put(45.,0.){\line(1,0){40.}}
\put(45.,50.){\line(1,0){40.}}
\put(50.,15.){\line(1,0){25.}}
\put(50.,35.){\line(1,0){25.}}
\put(80.,15.){\makebox(0,0)[cc]{$k$}}
\put(80.,35.){\makebox(0,0)[cc]{$j$}}
\put(62.5,15.){\oval(17.,17.)[l]}
\put(62.5,15.){\oval(17.,17.)[r]}
\put(60.,11.5){\vector(0,1){8.}}
\put(65.,18.5){\vector(0,-1){8.}}
\put(70.,25.){\oval(15.,15.)[r]}
\put(70,32.5){\vector(-1,0){0.1}}
\end{picture}$$
are also present. What the above indicates is that besides the number of
Cooper pairs being conserved in each system, the number of double pairs
(to be more precise, the number of energy levels which are completely
filled) is also conserved.  This can be seen
in each of the scattering processes depicted
graphically above. In each case the scattering does not overall change 
the number of completely filled levels. 
There are further symmetries in the Hamiltonian. For example, there is
a reflection symmetry which interchanges the labels 1 and 2 for the two
BCS systems. This arises as a result of a global $so(3)\oplus u(1)$
symmetry that the model possesses, which will be made more clear later.
In that which follows we shall first discuss
the integrability of the Hamiltonian (\ref{Ham}) 
in the context of the QISM.

In order to built up a mechanism to construct an integrable $su(4)$
pairing model, let us first recall the quantum $R$-matrix associated
with the Lie algebra $su(4)$, which acts in the tensor product of two
4-dimensional spaces $V \otimes V$ and can be written as 
\begin{equation}
R(\lambda )=\frac{\left(\lambda .I\otimes I 
+\eta P \right)}{\left(\lambda+\eta \right)}. 
\label{RM}
\end{equation}
Above $\lambda$ is the usual spectral parameter, $P$ is the 
permutation operator with matrix elements $ P_{\alpha\beta, \gamma \delta}=
\delta_{\alpha\delta}\delta_{\beta\gamma},~~\alpha,\beta, \gamma,
\delta=1,2,3,4 $ and $\eta $ is the quasiclassical limit parameter; i.e.
$$\lim_{\eta\rightarrow 0} R(\lambda)=I\otimes I.$$  
It is known that this $R$-matrix satisfies the Yang-Baxter
equation (YBE)
\begin{equation}
R_{12}(\lambda-\mu)R_{13}(\lambda)R_{23}(\mu)
=R_{23}(\mu)R_{13}(\lambda)R_{12}(\lambda-\mu). \label{YBE}
\end{equation}
The R-matrix
may be viewed as the structural constants for the Yang-Baxter
algebra generated by the monodromy matrix $T(\lambda)$, namely,
\begin{equation}
R_{12}(\lambda-\mu)\stackrel{1}{T}(\lambda)\stackrel{2}{T}(\mu)= 
\stackrel{2}{T}(\mu)\stackrel{1}{T}(\lambda)R_{12}(\lambda-\mu).\label{YBA}
\end{equation}
Consequently, the $R$-matrix (\ref{RM}) allows us to construct a 
realization of the monodromy matrix through
\begin{equation}
T(\lambda )=G_0R_{0\Omega}(\lambda -\epsilon_{\Omega })\cdots
G_0R_{01}(\lambda -\epsilon_1). \label{trans}
\end{equation}
Here the subscript $0$ denotes the auxiliary space and $G$
satisfying 
$$[R,\,G\otimes G]=0$$ 
is a class of c-valued solutions of
the YBE (\ref{YBE}).  As a consequence of the Yang-Baxter algebra (\ref{YBA}),
the transfer matrices $t(\lambda)=\rm{tr}_0T(\lambda)$ mutually commute for
different values of the spectral parameter $\lambda $. This transfer
matrix is the starting point in the construction of a $su(4)$-type
Gaudin Hamiltonian, from which we can obtain the $su(4)$ pairing
Hamiltonian, as will be shown below. For this purpose we make the following
identification for the basis states 
\begin{eqnarray}
& & |1\rangle=\,\,\,\,\,\,\,\,\,\,|0\rangle \,\,\,\,\, \,\,\,\,\,\,
=\,\,\,\,\,\,
\unitlength=0.50mm
\begin{picture}(20.,25.)
\put(5.,4.){\line(1,0){15.}}
\put(25.,4.){\line(1,0){15.}}
\end{picture} \nn \\
& &|2\rangle=b^+(1)b^+(2)|0\rangle =
\unitlength=0.50mm
\begin{picture}(20.,25.)
\put(5.,4.){\line(1,0){15.}}
\put(25.,4.){\line(1,0){15.}}
\put(31.,0.){\vector(0,1){10.}}
\put(35.,8.){\vector(0,-1){10.}}
\put(11.,0.){\vector(0,1){10.}}
\put(15.,8.){\vector(0,-1){10.}}
\end{picture}
\nn \\
& &
|3\rangle=\,\,\,\,\,\,b^+(1)|0\rangle \,\,\,\,\,\,=\,\,
\unitlength=0.50mm
\begin{picture}(20.,25.)
\put(5.,4.){\line(1,0){15.}}
\put(25.,4.){\line(1,0){15.}}
\put(11.,0.){\vector(0,1){10.}}
\put(15.,8.){\vector(0,-1){10.}}
\end{picture}
\nn \\
& &
|4\rangle=\,\,\,\,\,\,b^+(2)|0\rangle\,\,\,\,\,\,=\,\,
\unitlength=0.50mm
\begin{picture}(20.,25.)
\put(5.,4.){\line(1,0){15.}}
\put(25.,4.){\line(1,0){15.}}
\put(31.,0.){\vector(0,1){10.}}
\put(35.,8.){\vector(0,-1){10.}}
\end{picture}
\label{states} \end{eqnarray}
and choose the G-matrix to be given by
\begin{equation}
G\equiv \exp\left[\frac{2\eta (1-n(1)-n(2))}{\Omega g}\right]=
\left(\begin{array}{cccc}
\exp({\frac{2\eta}{\Omega g}})&0&0&0\\
0&\exp({\frac{-2\eta}{\Omega g }})&0&0\\
0&0&1&0\\
0&0&0&1
\end{array}\right),
\end{equation}
to construct the transfer matrix $t(\lambda)$. It  
 can be verified that 
\begin{eqnarray}
t(\epsilon _j) & = & G_jR_{j,j-1}(\epsilon_j-\epsilon_{j-1})\cdots G_jR_{j1}(\epsilon _j-\epsilon _1)G_jR_{j,\Omega}(\epsilon_j-\Omega )\nonumber\\
& &
\cdots G_jR_{j,j+1}(\epsilon_j-\epsilon_{j+1})G_j.
\end{eqnarray}
Next, taking the quasiclassical limit, we  find 
\begin{eqnarray}
R_{j,k}(\lambda )|_{\eta \rightarrow0}
&=&I\otimes I +\eta r_{j,k}(\lambda )+O(\eta ^2)\\
G_j|_{\eta \rightarrow 0} 
&=&I+\frac{2\eta }{\Omega g}\left(1-n_j(1)-n_j(2)\right) +O(\eta ^2) 
\end{eqnarray}
where $ r_{j,k}(\lambda)=\frac{P_{j,k}-1}{\lambda}$.
Thus it follows that
\begin{equation}
t(\epsilon _j)|_{\eta \rightarrow 0 }=1+\eta \left(\tau _j-\sum
  ^{\Omega}_{\mbox{\scriptsize $\begin{array}{c}k=1\\ k\neq
  j\end{array}$}} \frac{1}{\epsilon
  _j-\epsilon_k}\right)+\cdots
\end{equation}
where
\begin{equation}
\tau _j= \frac{2}{g}(1-n_j(1)-n_j(2))
+\sum^\Omega_{\mbox{\scriptsize $\begin{array}{c}k=1\\
  k\neq j\end{array}$}} \frac{\sum
  ^{4}_{\alpha,\beta}E^{\alpha\beta}_jE^{\beta
  \alpha}_k}{\epsilon_j-\epsilon _k}.\label{tau}
\end{equation}
Here $E^{\alpha \beta}=|\alpha \rangle\langle \beta|$, $\alpha,
\beta=1,\cdots,4$ are the Hubbard operators. An immediate consequence
from the Yang-Baxter algebra (\ref{YBA}) is that $\left[\tau _j,\tau
_k\right]=0$. In addition, as a result of the $so(3)\oplus u(1)$
symmetry mentioned earlier, it can be shown  that there are extra
conserved operators $K$ and $\chi$ such that   
$$\left[\tau_j,\,K\right]=\left[\tau_j,\,\chi\right]
=\left[K,\,\chi\right]=0. $$ 
Above,  $K$ is the Casimir operator of an  
$so(3)$ subalgebra acting on the $\Omega$-fold tensor product
\begin{equation}
K=\sum^{\Omega}_{j,k}\left(L_j^+L_k^-+L^-_jL^+_k
+\frac{1}{2}L_j^0L_k^0\right) \label{Casi}
\end{equation}
where ($L^0,\,L^+,\,L^-$) are the basis elements of this canonical
$so(3)$ subalgebra
\begin{eqnarray}
& &L^+=E^{34}=b^+(1)b(2),\nn \\
&&L^-=E^{43}=b^+(2)b(1),\nonumber\\
& &L^0=E^{33}-E^{44}=n(1)-n(2).\label{qn1}   
\end{eqnarray}
The $u(1)$ operator $\chi$ explicitly reads 
\bea \chi&=&\sum_{j=1}^\Omega (E^{33}_j+E^{44}_j) \nn \\
&=& \sum_{j=1}^\Omega \left(n_j(1)-n_j(2)\right)^2.  \label{qn2}  \eea 

Any Hamiltonian which is defined in terms of the mutually commuting
set of operators 
\begin{equation} 
\{\tau_j,\,K,\,\chi\} \label{conserved} \end{equation} 
will necessarily be integrable where the operators in (\ref{conserved})
represent the constants of the motion. By making the following choice 
\begin{eqnarray}
H & = & -g\sum ^{\Omega}_{j=1}\epsilon _j\tau
_j +\frac{g^3}{16}\sum_{j,k=1}^\Omega \tau_j\tau_k  
+\frac{3g^2}{4}\sum
^{\Omega}_{j=1}\tau
_j+\frac{g}{2}K
\nonumber\\
& & +\frac{g}{2}\chi\left(\chi-\Omega\right)+2\sum^{\Omega}_j\epsilon_j 
+\frac{g\Omega^2}{4}-2g\Omega  \label{H1}
\end{eqnarray}
we produce the Hamiltonian (\ref{Ham}). In order to determine the
energy spectrum of this model, we will need to determine the eigenvalues
of the conserved operators, to which we turn next.

\section{Bethe ansatz solutions}
\label{sec3}

Besides proving the integrability of the model, we can
also obtain its exact solution from the algebraic Bethe ansatz for
the standard $su(4)$ vertex model constructed from the R-matrix
(\ref{RM}). Employing the nested  algebraic Bethe ansatz \cite{MR} we
can obtain the eigenvalue of the transfer matrix (\ref{trans}) as
\begin{eqnarray}
\Lambda (v,\{\lambda_j\}\{u_i \})
&=&e^{\frac{2\eta }{g}}\prod^{N}_{i=1}\frac{v-v_i-\eta}{v-v_i}\nonumber\\
& &
+e^{-\frac{2\eta }{g}}\prod_{i=1}^{\Omega}\frac{v-\epsilon_i-\frac{\eta}{2}}{v-\epsilon_i+\frac{\eta}{2}}\prod_{i=1}^{N}\frac{v-v_i+\eta}{v-v_i}\prod_{l=1}^{M}\frac{v-u_l-\frac{\eta}{2}}{v-u_l+\frac{\eta}{2}}\nonumber\\
& &
+\prod_{i=1}^{\Omega}\frac{v-\epsilon_i-\frac{\eta}{2}}{v-\epsilon_i+\frac{\eta}{2}}\prod_{l=1}^{M}\frac{v-u_l+\frac{3\eta}{2}}{v-u_l+\frac{\eta}{2}}
\prod^{Q}_{j=1}\frac{v-w_j}{v-w_j+\eta}\nonumber\\
& &
+\prod_{i=1}^{\Omega}\frac{v-\epsilon_i-\frac{\eta}{2}}{v-\epsilon_i+\frac{\eta}{2}}\prod^{Q}_{j=1}\frac{v-w_j+2\eta}{v-w_j+\eta}.\label{Teigen}
\end{eqnarray}
Above the parameters $v_j , u_m$ and $w_k$ satisfy the Bethe ansatz equations
\begin{eqnarray}
& &e^{-\frac{4\eta}{g}}\prod ^{\Omega}_{i=1}\frac{v _j-\epsilon _i-\frac{\eta}{2}}{v_j-\epsilon _i+\frac{\eta}{2}}
\prod^{N}_{\stackrel{\scriptstyle l=1}{l\neq j}}
\frac{v_j-v_l+\eta}{v_j-v_l-\eta}\prod^{M}_{l=1}\frac{v_j-u_l-\frac{\eta}{2}}
{v_j-u_l+\frac{\eta}{2}}=1 \nn \\
& &
e^{-\frac{2\eta}{g}}\prod^{N}_{i=1}\frac{u_m-v_i+\frac{\eta}{2}}{u_m-v_i-\frac{\eta}{2}}=
\prod^{M}_{\stackrel{\scriptstyle i=1}{i\neq m}}\frac{u_m-u_i+\eta}{u_m-u_i-\eta}
\prod^{Q}_{l=1}\frac{u_m-w_l-\frac{\eta}{2}}{u_m-w_l+\frac{\eta}{2}}\nn
\\
&& 
\prod^{M}_{l=1}\frac{w_k-u_l+\frac{\eta}{2}}{w_k-u_l-\frac{\eta}{2}}
=\prod^{Q}_{\stackrel{\scriptstyle l=1}{l\neq k}}\frac{w_k-w_l+\eta}
{w_k-w_l-\eta} \nn   \\
& &j=1,\cdots,N,~~~m=1,\cdots,M,~~~k=1,\cdots,Q.\nonumber
\end{eqnarray}
Defining   $N(a)=\sum_{j=1}^\Omega n_j(a)$,  we can readily determine that 
the quantum numbers $N,\,M$ and $Q$ are given by 
\bea N&=&N(1)+N(2)-N(1)N(2), \nn \\ 
M&=&N(1)+N(2)-2N(1)N(2), \nn \\ 
Q&=& N(2)-N(1)N(2). \label{qn3}  \eea   

The
eigenvalues of the integrals of motion $\tau _j$ (\ref{tau}) can be
obtained from the expansion of the eigenvalue of the transfer matrix
(\ref{Teigen}) in the parameter $\eta$. Explicitly, the eigenvalues of
$\tau _j$ are given by
\begin{equation}
\Lambda _j=\frac{2}{g}+\sum^{N}_{l=1}\frac{1}{v _l-\epsilon _j}+\sum ^{\Omega }_{\stackrel{\scriptstyle k=1}{k\neq j}}\frac{1}{\epsilon _j-\epsilon_k},\label{eigen-tau}
\end{equation}
where the parameters  satisfy  the following  equations
\begin{eqnarray}
& &
\frac{4}{g}+\sum^{\Omega}_{i=1}\frac{1}{v_j-\epsilon_i}+
\sum_{l=1}^{M}\frac{1}{v_j-u_l}=
2\sum^N_{\stackrel{\scriptstyle l=1}{l\neq j}}\frac{1}{v_j-v_l},
\nn \\
& &
\frac{2}{g}-\sum^{N}_{i=1}\frac{1}{u_m-v_i}-2\sum^M_{\stackrel{\scriptstyle l=1}
{l\neq m}}\frac{1}{u_l-u_m}=\sum^{Q}_{l=1}\frac{1}{u_m-w_l},\nn \\
& &
\sum^{M}_{l=1}\frac{1}{w_k-u_l}
=2\sum^{Q}_{\stackrel{\scriptstyle l=1}{l\neq k}}\frac{1}{w_k-w_l},
\label{bae}   \\
& &j=1,\cdots,N,~~~m=1,\cdots,M,~~~k=1,\cdots,Q \nonumber
\end{eqnarray}
We will also need the eigenvalues of the operators $K$ and $\chi$. 
Through use of (\ref{qn1}, \ref{qn2}, \ref{qn3}) we find that $\chi$ has eigenvalue
$M$ while the eigenvalues of $K$ are 
$$ \frac{1}{2}(M-2Q)(M-2Q+2). $$ 

Finally, utilizing  (\ref{eigen-tau}) and noting the following identities
which can be derived from (\ref{bae}):
\begin{eqnarray}
& &
\sum^{Q}_{k=1}\sum^{M}_{l=1}\frac{1}{w_k-u_l}=2\sum^{Q}_{k=1}\sum^{Q}_{l=1}\frac{1}{w_k-w_l}=0,\nonumber \\
& &\sum^{M}_{m=1}\sum^{N}_{i=1}\frac{1}{u_m-v_i}=\frac{2M}{g},\nonumber\\
& &
\sum^{N}_{j=1}\sum^{\Omega}_{i=1}\frac{1}{v_j-\epsilon_i}=-\frac{2(2N-M)}{g},\nonumber\\
& &
\sum^{Q}_{k=1}\sum^{M}_{l=1}\frac{u_l}{u_l-w_k}-\sum^{Q}_{k=1}\sum^{M}_{l=1}\frac{w_k}{u_l-w_k}=MQ,\nonumber\\
& &
\sum^{M}_{m=1}\sum^{N}_{i=1}\frac{v_i}{v_i-u_m}-\sum^{M}_{m=1}\sum^{N}_{i=1}\frac{u_m}{v_i-u_m} =MN,\nonumber\\
& &
\sum^{N}_{j=1}\sum^{\Omega}_{i=1}\frac{\epsilon_i}{\epsilon_i-v_j}-\sum^{N}_{j=1}\sum^{\Omega}_{i=1}\frac{v_j}{\epsilon_i-v_j}=N\Omega,\nonumber\\
& &
\sum^{Q}_{k=1}\sum^{M}_{l=1}\frac{w_k}{w_k-u_l}=Q(Q-1),\nonumber\\
& &
-\frac{2}{g}\sum^M_{m=1}u_m+\sum^{M}_{m=1}\sum^{N}_{i=1}\frac{u_m}{u_m-v_i}+
\sum^{M}_{m=1}\sum^{Q}_{l=1}\frac{u_m}{u_m-w_l}=M(M-1),\nonumber\\
& &
\frac{4}{g}\sum^{N}_{j=1}v_j-\sum^{N}_{j=1}\sum^{\Omega}_{i=1}\frac{v_j}{\epsilon_i-v_j}+\sum^{N}_{j=1}\sum^{M}_{l=1}\frac{v_j}{v_j-u_l}=N(N-1),\nonumber
\end{eqnarray}
we can present from the relation (\ref{H1}) the eigenvalue of the Hamiltonian
(\ref{Ham}) as 
\begin{eqnarray}
E & = & 4\sum^{N}_{i=1}v_i-2\sum^M_{m=1}u_m
-g(2N-3M).
\end{eqnarray}
Let us make some small remarks about the degeneracies of the spectrum.
Though the eigenstates of the Hamiltonian have not been made explicit
here, it can be deduced by the standard arguments (e.g. \cite{hw})
that each is a
highest weight state with respect to the $so(3)$ symmetry algebra
(\ref{qn1}). In particular, the highest weight which is given by the
eigenvalue of the operator $L^0$ is $M-2Q$, so we can conclude that 
the multiplet generated by (\ref{qn1}) acting on this highest weight
state has dimension $M-2Q+1$. Therefore for each solution of (\ref{bae})  
with  given $N,\,M$ and $Q$,  the corresponding energy level has degeneracy
$M-2Q+1$. 
 
\section{Asymptotic solutions}
\label{sec4}

As the Bethe ansatz equations (\ref{bae}) take the form of coupled
non-linear equations it is unlikely to find analytic solutions, and one
tends to resort to numerical analysis. It is  however possible to
conduct an asymptotic analysis for small values of the coupling
parameter $g$. Below we undertake this for the ground state of the
system and some elementary excitations.  

For the ground state, we consider first the case $g=0$. Letting 
$N_1$ and $N_2$ denote the number of Cooper pairs in each system, then
it is clear that the ground state corresponds to filling the Fermi sea,
which is illustrated in Fig. (A) below, where without loss of generality
we assume that $N_1>N_2$.
For small $g\neq 0$ we see that the ground state will be described by a
solution of (\ref{bae}) with 
$N=N_1,\,M=N_1-N_2,\,Q=0$.  
$$
\unitlength=0.50mm
\begin{picture}(100.,85.)
\put(20.,76.){\makebox(0,0)[cc]{$(1)$}}
\put(0.,0.){\line(0,1){70.}}
\put(40.,0.){\line(0,1){70.}}
\put(0.,0.){\line(1,0){40.}}
\put(0.,70.){\line(1,0){40.}}
\put(5.,10.){\line(1,0){20.}}
\put(5.,25.){\line(1,0){20.}}
\put(5.,40.){\line(1,0){20.}}
\put(5.,65.){\line(1,0){20.}}
\put(5.,55.){\line(1,0){20.}}
\put(30.,55.){\makebox(0,0)[cc]{$N_1$}}
\put(17.5,55.){\oval(10.,10.)[l]}
\put(17.5,55.){\oval(10.,10.)[r]}
\put(15.,52.5){\vector(0,1){5.}}
\put(20.,57.5){\vector(0,-1){5.}}
\put(17.5,40.){\oval(10.,10.)[l]}
\put(17.5,40.){\oval(10.,10.)[r]}
\put(15.,37.5){\vector(0,1){5.}}
\put(20.,42.5){\vector(0,-1){5.}}
\put(17.5,10.){\oval(10.,10.)[l]}
\put(17.5,10.){\oval(10.,10.)[r]}
\put(15.,7.5){\vector(0,1){5.}}
\put(20.,12.5){\vector(0,-1){5.}}
\put(17.5,25.){\oval(10.,10.)[l]}
\put(17.5,25.){\oval(10.,10.)[r]}
\put(15.,22.5){\vector(0,1){5.}}
\put(20.,27.5){\vector(0,-1){5.}}
\put(65.,76.){\makebox(0,0)[cc]{$(2)$}}
\put(45.,0.){\line(0,1){70.}}
\put(85.,0.){\line(0,1){70.}}
\put(45.,0.){\line(1,0){40.}}
\put(45.,70.){\line(1,0){40.}}
\put(50.,10.){\line(1,0){20.}}
\put(50.,25.){\line(1,0){20.}}
\put(50.,40.){\line(1,0){20.}}
\put(50.,55.){\line(1,0){20.}}
\put(50.,65.){\line(1,0){20.}}
\put(75.,25.){\makebox(0,0)[cc]{$ N_2$}}
\put(62.5,10.){\oval(10.,10.)[l]}
\put(62.5,10.){\oval(10.,10.)[r]}
\put(60.,7.5){\vector(0,1){5.}}
\put(65.,12.5){\vector(0,-1){5.}}
\put(62.5,25.){\oval(10.,10.)[l]}
\put(62.5,25.){\oval(10.,10.)[r]}
\put(60.,22.5){\vector(0,1){5.}}
\put(65.,27.5){\vector(0,-1){5.}}
\put(39.,-7.){\makebox(0,0)[cc]{${\rm Fig. (A)}$}}
\end{picture}
\vspace{0.7cm}
$$
Thus the Bethe equations (\ref{bae}) 
reduce to two levels for the parameters $v_j$ and
$u_m$. For a small $g>0$ it is appropriate to consider the
asymptotic solution
\begin{eqnarray}
v_j&=&\epsilon _j+g\delta _j+g^2\sigma_j,~~~~~~~~~~~~\,\,\,
j=1,\cdots,N_1,\label{app}\\
u_m&=&\epsilon _{N_1-m+1}+g\alpha_m +g^2\beta _m,\,\,\,m=1,\cdots,N_1-N_2.
\end{eqnarray}
Substituting these  into the  Bethe equations (\ref{bae}) 
with the  configuration $N=N_1,\,M=N_1-N_2,\,Q=0$,
 one can find that 
\begin{eqnarray}
v_j & \approx & \epsilon_j-\frac{g}{4}+\frac{g^2}{16}\left[\sum^{\Omega}_{i=N_1+1}\frac{1}{\epsilon_j-\epsilon_i}-\sum^{N_2}_{\stackrel{\scriptstyle l=1}{l\neq j}}\frac{1}{\epsilon_j-\epsilon_l}\right],\,\,j\leq N_2, \label{asymp1}\\
v_{j} & \approx & \epsilon_{j}-\frac{g}{2}+\frac{g^2}{4}\left[\sum^{\Omega}_{i=N_1+1}\frac{1}{\epsilon_{j}-\epsilon_i}-\sum^{N_1}_{\stackrel{\scriptstyle l=N_2+1}{l\neq j}}\frac{1}{\epsilon_j-\epsilon_l}\right],\,\,j>N_2,\label{asymp2}\\
u_{m} & \approx & \epsilon_{N_1-m+1}+\frac{g^2}{4}\left[\sum^{\Omega}_{i=N_1+1}\frac{1}{\epsilon_{N_1-m+1}-\epsilon_i}+\sum^{N_2}_{\stackrel{\scriptstyle i=1}{i\neq N_1-m+1}}\frac{1}{\epsilon_{N_1-m+1}-\epsilon_i}\right.\nonumber\\
& &\left.~~~~~~-\sum^{N_1}_{\stackrel{\scriptstyle l=N_2+1}{l\neq N_1-m+1}}
\frac{2}{\epsilon_{N_1-m+1}-\epsilon_l}\right],\,\,
m=1,\cdots,N_1-N_2.\label{asymp3}
\end{eqnarray}
\vfil\eject 
The asymptotic ground state energy is  deduced to be given  by 
\begin{eqnarray}
E_0& \approx &
4\sum^{N_1}_{j=1}\epsilon_j -2\sum^{N_1}_{l=N_2+1}\epsilon_l
-(N_1+2N_2)g\nonumber\\
& &
+\frac{g^2}{4}\left[\sum^{N_2}_{j=1}\sum^{\Omega}_{i=N_1+1}
\frac{1}{\epsilon_j-\epsilon_i}
+\sum_{j=N_2+1}^{N_1}\left(\sum^{\Omega}_{i=N_1+1}
\frac{2}{\epsilon_{j}-\epsilon_i}-\sum^{N_2}_{i=1}
\frac{2}{\epsilon_{j}-\epsilon_i}\right)\right]. \nn 
\end{eqnarray}
It is important to point out that from the above ground state energy we
can infer some results about the asymptotic behaviour of zero
temperature correlation functions. Specifically, by employing the
Hellmann-Feynman theorem we have that 
$$\left<n_i(1)+n_i(2)\right>=\frac{1}{2}
\frac{\partial E_0}{\partial \epsilon_i}. $$
The result obtained is 
\bea 
\left<n_i(1)+n_i(2)\right>&\approx& \frac{g^2}{8}\left(\sum_{j=1}^{N_2} 
\frac{1}{(\epsilon_j-\epsilon_i)^2}+\sum_{j=N_2+1}^{N_1}\frac{2}{(\epsilon_j
-\epsilon_i)^2}\right)~~~~~~{\rm for\,}\,i>N_1, \nn \\
\left<n_i(1)+n_i(2)\right>&\approx&1- \frac{g^2}{8}\left(\sum_{j=N_1+1}^\Omega 
\frac{2}{(\epsilon_j-\epsilon_i)^2}-\sum_{j=1}^{N_2}\frac{2}{(\epsilon_j 
-\epsilon_i)^2}\right)~{\rm for\,}\,N_1\geq i>N_2, \nn \\
\left<n_i(1)+n_i(2)\right>&\approx&2- \frac{g^2}{8}\left(\sum_{j=N_1+1}^\Omega
\frac{1}{(\epsilon_j-\epsilon_i)^2}+\sum_{j=N_2+1}^{N_1}\frac{2}{(\epsilon_j
-\epsilon_i)^2}\right)~{\rm for\,}\,i\leq N_2  . \nn \ 
\eea 

Next let us consider a possible excitation which can be obtained by
breaking one Cooper pair in BCS(2). In the $g=0$ case, the excited state is
depicted in Fig. (B). 
$$
\unitlength=0.50mm
\begin{picture}(100.,85.)
\put(20.,76.){\makebox(0,0)[cc]{$(1)$}}
\put(0.,0.){\line(0,1){70.}}
\put(40.,0.){\line(0,1){70.}}
\put(0.,0.){\line(1,0){40.}}
\put(0.,70.){\line(1,0){40.}}
\put(5.,10.){\line(1,0){20.}}
\put(5.,25.){\line(1,0){20.}}
\put(5.,40.){\line(1,0){20.}}
\put(5.,65.){\line(1,0){20.}}
\put(5.,55.){\line(1,0){20.}}
\put(30.,55.){\makebox(0,0)[cc]{$N_1$}}
\put(17.5,55.){\oval(10.,10.)[l]}
\put(17.5,55.){\oval(10.,10.)[r]}
\put(15.,52.5){\vector(0,1){5.}}
\put(20.,57.5){\vector(0,-1){5.}}
\put(17.5,40.){\oval(10.,10.)[l]}
\put(17.5,40.){\oval(10.,10.)[r]}
\put(15.,37.5){\vector(0,1){5.}}
\put(20.,42.5){\vector(0,-1){5.}}
\put(17.5,10.){\oval(10.,10.)[l]}
\put(17.5,10.){\oval(10.,10.)[r]}
\put(15.,7.5){\vector(0,1){5.}}
\put(20.,12.5){\vector(0,-1){5.}}
\put(17.5,25.){\oval(10.,10.)[l]}
\put(17.5,25.){\oval(10.,10.)[r]}
\put(15.,22.5){\vector(0,1){5.}}
\put(20.,27.5){\vector(0,-1){5.}}
\put(65.,76.){\makebox(0,0)[cc]{$(2)$}}
\put(45.,0.){\line(0,1){70.}}
\put(85.,0.){\line(0,1){70.}}
\put(45.,0.){\line(1,0){40.}}
\put(45.,70.){\line(1,0){40.}}
\put(50.,10.){\line(1,0){20.}}
\put(50.,25.){\line(1,0){20.}}
\put(50.,40.){\line(1,0){20.}}
\put(50.,55.){\line(1,0){20.}}
\put(50.,65.){\line(1,0){20.}}
\put(75.,25.){\makebox(0,0)[cc]{$ N_2$}}
\put(62.5,10.){\oval(10.,10.)[l]}
\put(62.5,10.){\oval(10.,10.)[r]}
\put(60.,7.5){\vector(0,1){5.}}
\put(65.,12.5){\vector(0,-1){5.}}
\put(62.5,22.5){\vector(0,1){7.}}
\put(62.5,42.5){\vector(0,-1){7.}}
\put(39.,-7.){\makebox(0,0)[cc]{${\rm Fig. (B)}$}}
\end{picture}
\vspace{0.7cm}
$$
For non-zero $g$, we choose
  $N=N_1-2,\,M=N_1-N_2-1$ and block the levels with energy
  $\epsilon_{N_2},\,\epsilon_{N_2+1}$.
 {}From the asymptotic solutions (\ref{asymp1}),
 (\ref{asymp2}) and (\ref{asymp3}),
we obtain the excitation energy
\begin{eqnarray}
E_1& \approx &4\sum^{N_1}_{j=1}\epsilon_j 
-2\sum^{N_1}_{l=N_2+2}\epsilon_l-\epsilon_{N_2}-\epsilon_{N_2+1}
-(N_1+2N_2-4)g\nonumber\\
& &
+\frac{g^2}{4}\left[\sum^{N_2-1}_{j=1}\sum^{\Omega}_{i=N_1+1}\frac{1}{\epsilon_j-\epsilon_i}
+\sum_{j=N_2+2}^{N_1}\left(\sum^{\Omega}_{i=N_1+1}\frac{2}
{\epsilon_{j}-\epsilon_i}-\sum^{N_2-1}_{i=1}\frac{2}
{\epsilon_{j}-\epsilon_i}\right)\right].\nn 
\end{eqnarray}
Therefore the  gap obtained  through the breaking a Cooper pair in 
BCS(2)
is  given by 
\bea 
\Delta _1 &\approx& \epsilon_{N_2+1}-\epsilon_{N_2}+4g
+\frac{g^2}{4}\left[\sum^{\Omega}_{i=N_1+1}
\frac{1}{\epsilon_{i}-\epsilon_{N_2}}
+\sum_{i=N_1+1}^\Omega\frac{2}{\epsilon_i-\epsilon_{N_2+1}} 
\right. \nn \\ 
&&~~~~~~~~~~~~~
~~~~~~~~~~~\left.+\sum_{j=N_2+2}^{N_1}\frac{2}{\epsilon_j-\epsilon_{N_2}}
+\sum_{i=1}^{N_2}\frac{2}{\epsilon_{N_2+1}-\epsilon_i}\right] \nn \eea  

Another possibility for an 
excitation at $g=0$ is to break the Cooper pair at level
$\epsilon_{N_1}$ in  BCS(1) (see Fig.(C)). For $g\neq 0$, 
the configuration should be  accommodated as
 $N=N_1-1,\,M=N_1-N_2-1$  with the levels $\epsilon_{N_1},\,
 \epsilon_{N_1+1} $ blocked.  
$$
\unitlength=0.50mm
\begin{picture}(100.,85.)
\put(20.,76.){\makebox(0,0)[cc]{$(1)$}}
\put(0.,0.){\line(0,1){70.}}
\put(40.,0.){\line(0,1){70.}}
\put(0.,0.){\line(1,0){40.}}
\put(0.,70.){\line(1,0){40.}}
\put(5.,10.){\line(1,0){20.}}
\put(5.,25.){\line(1,0){20.}}
\put(5.,40.){\line(1,0){20.}}
\put(5.,65.){\line(1,0){20.}}
\put(5.,55.){\line(1,0){20.}}
\put(30.,55.){\makebox(0,0)[cc]{$N_1$}}
\put(17.5,52.5){\vector(0,1){7.}}
\put(17.5,67.5){\vector(0,-1){7.}}
\put(17.5,40.){\oval(10.,10.)[l]}
\put(17.5,40.){\oval(10.,10.)[r]}
\put(15.,37.5){\vector(0,1){5.}}
\put(20.,42.5){\vector(0,-1){5.}}
\put(17.5,10.){\oval(10.,10.)[l]}
\put(17.5,10.){\oval(10.,10.)[r]}
\put(15.,7.5){\vector(0,1){5.}}
\put(20.,12.5){\vector(0,-1){5.}}
\put(17.5,25.){\oval(10.,10.)[l]}
\put(17.5,25.){\oval(10.,10.)[r]}
\put(15.,22.5){\vector(0,1){5.}}
\put(20.,27.5){\vector(0,-1){5.}}
\put(65.,76.){\makebox(0,0)[cc]{$(2)$}}
\put(45.,0.){\line(0,1){70.}}
\put(85.,0.){\line(0,1){70.}}
\put(45.,0.){\line(1,0){40.}}
\put(45.,70.){\line(1,0){40.}}
\put(50.,10.){\line(1,0){20.}}
\put(50.,25.){\line(1,0){20.}}
\put(50.,40.){\line(1,0){20.}}
\put(50.,55.){\line(1,0){20.}}
\put(50.,65.){\line(1,0){20.}}
\put(75.,25.){\makebox(0,0)[cc]{$ N_2$}}
\put(62.5,10.){\oval(10.,10.)[l]}
\put(62.5,10.){\oval(10.,10.)[r]}
\put(60.,7.5){\vector(0,1){5.}}
\put(65.,12.5){\vector(0,-1){5.}}
\put(62.5,25.){\oval(10.,10.)[l]}
\put(62.5,25.){\oval(10.,10.)[r]}
\put(60.,22.5){\vector(0,1){5.}}
\put(65.,27.5){\vector(0,-1){5.}}
\put(39.,-7.){\makebox(0,0)[cc]{${\rm Fig. (C)}$}}
\end{picture}
\vspace{0.7cm}
$$
We find the gap obtained in this case 
to be given by 
\bea 
\Delta _2&\approx& \epsilon_{N_1+1}-\epsilon_{N_1}+g
+\frac{g^2}{4}\left[\sum^{N_2}_{ j=1}
\frac{1}{\epsilon_{N_1+1}-\epsilon_{j}}
+\sum^{N_2}_{i=1}
\frac{2}{\epsilon_{N_1}-\epsilon_{i}}
\right. \nn   \\ 
&&~~~~~~~~~~~~~~~~~~~~~~~~~
\left.+\sum_{j=N_2+1}^{N_1-1}\frac{2}{\epsilon_{N_1+1}-\epsilon_j}+ 
\sum^{\Omega}_{i=N_1+1}\frac{2}{\epsilon_i-\epsilon_{N_1}}\right].
\nn \eea  

For the asymptotics to be valid we require from 
(\ref{asymp1}, \ref{asymp2}, \ref{asymp3}) that
$g<<\epsilon_i$ for all $i$. However in the above two expressions for the
energy gaps  the value of $g$ may still be larger than the spacings 
$\epsilon_{i+1}-\epsilon_i$ of
the single particle energies. 
So we can see from the analysis when this is the case that the inclusion of
the pairing interactions does produce a significant gap between the
ground state and the first excited states.

\section{Conclusion}
\label{sec5}

To summarize, we have constructed an integrable pairing Hamiltonian
based on the $su(4)$ Lie algebra. This model can be interpreted as
describing two coupled BCS systems of different types, such as for
protons and neutrons in a nuclear system. The
Bethe ansatz equations and the energies of the model have been
calculated. For small values of the coupling parameter $g$, 
we asymptotically analyzed the ground state and
elementary excitations, and the expectation values for the occupation
numbers. 
An open problem that we will address in the future is 
the exact calculation of form factors and correlation functions using
the techniques developed by Babujian et al. in \cite{Karowski}.
  
{\bf Acknowledgments} 
XWG gratefully acknowledges the Centre for Mathematical Physics, the
University of the Queensland for kind hospitality and FAPERGS
(Funda\c{c}\~{a}o de Amparo \~{a} Pesquisa do Estado do Rio Grande do
Sul) for financial support. AF would like to acknowledge M. Karowski
and H. Babujian for discussions and the group of Theoretical Physics
of the FU-Berlin for their kind hospitality. She also thanks
DAAD(Deutsche Akademischer Austauschdienst) and FAPERGS for 
financial support.  HQZ and JL thank the Australian Research
Council.


\end{document}